\documentclass[12pt,english]{article}
\usepackage[T1]{fontenc}
\usepackage[latin9]{inputenc}
\usepackage[letterpaper]{geometry}
\usepackage{amsmath}
\usepackage{graphicx}

\makeatletter
\newcommand{\lyxaddress}[1]{
	\par {\raggedright #1
	\vspace{1.4em}
	\noindent\par}
}

\usepackage{times}

\makeatother

\usepackage{babel}
\usepackage[sorting=none, maxbibnames=99]{biblatex}
\begin{document}
\title{Analytic Solution for the Linear Rheology of Living Polymers}
\author{Vickie Chen$^{1}$, Charles T. Drucker$^{1}$, Claire Love$^{1}$, Jonathon Peterson$^{2}$, and Joseph D. Peterson$^{1}$}
\maketitle

\lyxaddress{$^{1}$University of California Los Angeles, Department of Chemical and Biomolecular Engineering, 420 Westwood Pl, Los Angeles CA 90095, $^{2}$Purdue University, Department of Mathematics, 150N University Street, West Lafayette, IN 47907}
\begin{abstract}
It is often said that well-entangled and fast-breaking living polymers (such as wormlike micelles) exhibit a single relaxation time in their reptation dynamics, but the full story is somewhat more complicated. Understanding departures from single-Maxwell behavior is crucial for fitting and interpreting experimental data, but in some limiting cases numerical methods of solving living polymer models can struggle to produce reliable predictions/interpretations. In this work, we develop an analytic solution for the shuffling model of living polymers. The analytic solution is a converging infinite series, and it converges fastest in the fast-breaking limit where other methods can struggle.
\end{abstract}

\section{Introduction and Background}

\global\long\def\tb{\tau_{\text{B}}}%

\global\long\def\trep{\tau_{\text{rep}}}%

\global\long\def\ppt{\frac{\partial}{\partial t}}%

\global\long\def\ddt{\frac{d}{dt}}%

\global\long\def\lmin{L_{\text{min}}}%

\global\long\def\lmax{L_{\text{max}}}%

\global\long\def\tmin{\tau_{\text{min}}}%

\global\long\def\kmax{k_{\text{max}}}%

\global\long\def\zb{\zeta_{B}}%

The fundamental theory of stress relaxation in well-entangled living polymers was first put forward more than three decades ago by Cates \cite{cates1987reptation}. To summarize Cates' theory, well-entangled living polymers are constrained to relax their stress via reptation in a tube \cite{doi1978dynamics} but permitted to break along their contours and attach at their ends. The reversible scission reactions yield no net change to the molecular weight distribution (which is assumed to be at equilibrium) but they do reorganize tube segments - interior segments (which relax slowly) become end segments when a break occurs, and end segments (which are already relaxed) move to the interior with every recombination.

In the limit where polymers do not break very quickly compared to the typical time for reptation to occur, the effect of living polymer reactions can be neglected for the purpose of rheology \cite{cates1987reptation,peterson2021constitutive}. In the opposite limit, where reversible scission is much faster than reptation, there are two major changes to the material's rheological properties. First, stress relaxation occurs on faster timescales than would otherwise be possible - interior segments are always being shuffled to end positions where they can relax more quickly. Second, stress relaxation primarily occurs with a single characteristic timescale - for any interior tube segment, the rate limiting step for stress relaxation is the same (waiting to become an end segment) \cite{cates1987reptation}.

To rephrase the preceding paragraph in a more quantitative way, Cates showed that if the typical time for a polymer to break, $\tb$, is much faster than the typical time for reptation, $\trep$, the complex modulus $G^{*}(\omega)$ will exhibit ideal Maxwell behavior:

\begin{equation}
\frac{G^{*}(\omega)}{G_{e}}=\frac{i\omega\tau}{1+i\omega\tau}\hspace{1cm}\text{for \ensuremath{\omega\tb\ll1} and \ensuremath{\tb/\trep\ll1}}\label{eq:single_maxwell}
\end{equation}
where $\tau$ is the Maxwell relaxation time and $G_{e}$ is a shear modulus for the material. Cates further showed that in this same ``fast breaking'' limit, $\zb=\tb/\trep\ll1$, the relaxation time $\tau$ has an asymptotic scaling $\tau\sim[\tb\trep]^{1/2})$.

These results have proven useful for qualitatively interpreting experimental observations in wormlike micelles and other living polymer systems \cite{cates1990statics,kern1991rheological,vereroudakis2021tunable,louhichi2017humidity}, but for quantitative interpretation, equation \ref{eq:single_maxwell} does not provide enough information to uniquely determine $\tb$ and $\trep$ from the composite relaxation time $\tau$. Sometimes - erroneously - a value of $\tb$ is inferred from a local minima in the loss modulus \cite{rogers2014rheology,kim2013superposition,louhichi2017humidity}, but this correlation has no real basis in theory to the best of our knowledge \cite{granek1994dip}. Fortunately, equation \ref{eq:single_maxwell} is incomplete (only valid for $\omega\tb\ll1$) and a more complete description of living polymer rheology is possible.

A practical means of assessing $G^{*}(\omega)$ across all frequencies - and uniquely specifying $\tb$ and $\trep$ - was put forward by Granek and Cates in their landmark Poisson Renewal model \cite{granek1992stress}. Unlike equation \ref{eq:single_maxwell}, however, the Poisson renewal model did not admit a closed-form solution and had to be solved numerically. However, the Poisson renewal model is so simple to solve numerically that (in our opinion) there is no serious engineering need for an analytic solution.

In the years since the Poisson renewal model was published, there have been two significant developments:

First, there has been an expansion in our understanding of polymer physics and a corresponding expansion of living polymer modeling tools \cite{zou2014mesoscopic,zou2015determination,tan2021determining,scattering1,branchML,sato2020slip,peterson2020full,peterson2021constitutive,pahari2021slip,Love2024Numerical}. This proliferation of models has lead to some uncertainty regarding which models should be preferred for interpreting experimental data - however a recent review has shown that many of these models have a similar ability to fit experimental data, albeit with slightly different quantitative interpretations of the fit parameters \cite{peterson2023wormlike}. In many cases, the best model to use will be whichever model is easiest to implement.

Second, a closer inspection of the assumptions and approximations of the Poisson Renewal model has suggested that the simplest possible implementation of Poisson renewal (length-independent renewal time) connects with a better physical interpretation of the true reversible scission process \cite{peterson2020full,peterson2023wormlike}. This reinterpretation of the Poisson renewal process has been called the shuffling model.

Together, these developments suggest that an analytic solution - analagous to equation \ref{eq:single_maxwell} but valid for all $\zb$ and all $\omega$ - may be both preferrable and possible. In this manuscript, we will provide such a result; section \ref{sec:Derivation} contains a derivation culminating in an infinite series solution (equation \ref{eq:solution}), and section \ref{sec:Validation} validates that result by comparing against converged calculations obtained via traditional quadrature methods.

\section{\label{sec:Derivation}Derivation}

The shuffling model for reptation in living polymers describes the tube survival probability $P$ as a function of time $t$, chain length $L$, and chain contour position $s$. Polymers are assumed to relax their stress by reptation, and tube sections are randomly shuffled through the system on a timescale $\tb$ \cite{peterson2020full}. The model assumes an equilbrium, exponential molecular weight distribution $n(L)\sim\exp(-L/\bar{L})$ and curviliear diffusion constant for reptation that scales inversely with chain length, $D_{C}/L$:

\begin{equation}
\ppt P=\frac{D_{C}}{L}\frac{\partial^{2}P}{\partial s^{2}}-\frac{1}{\tb}(P-\bar{P})\label{eq:shuffling}
\end{equation}

\begin{equation}
\bar{P}=\frac{\int_{0}^{\infty}dLe^{-L/\bar{L}}\int_{0}^{L}dsP(t,s,L)}{\int_{0}^{\infty}dLLe^{-L/\bar{L}}}
\end{equation}

\begin{equation}
P(t,s=0,L)=P(t,s=L,L)=0\hspace{1cm}P(t=0,s,L)=1
\end{equation}

The tube survival probability $P$ is related to the complex modulus $G^{*}(\omega)$ via a one-side Fourier transform:

\begin{equation}
\frac{G^{*}(\omega)}{G_{e}}=i\omega\int_{0}^{\infty}dte^{-i\omega t}\bar{P}(t)
\end{equation}

By applying the one-side Fourier transform step to equation \ref{eq:shuffling} directly, one can obtain a simplified expression for the complex modulus $G^{*}(\omega)$ \cite{granek1992stress,peterson2020full,peterson2023wormlike}:

\begin{equation}
\frac{G^{*}(\omega)}{G_{e}}=i\omega\trep\left[\frac{1}{C(\zb,\omega)}-\frac{1}{\zb}\right]^{-1}
\end{equation}

\begin{equation}
C(\zb,\omega)=\int_{0}^{\infty}dz\frac{ze^{-z}}{\beta}\left[1-\frac{1}{\alpha}\tanh\left(\alpha\right)\right]\label{eq:C_eqn}
\end{equation}

\begin{equation}
\alpha=\frac{z^{3/2}\beta^{1/2}}{2}\hspace{1cm}\beta=\zb^{-1}+i\omega\trep
\end{equation}

The above results are all well known, but until now an analytic solution to equation \ref{eq:C_eqn} has not been produced. To proceed towards an analytic solution, we first expand the $\tanh(\alpha)$ term \ref{eq:C_eqn} using $\tanh(\alpha)=1+2\sum_{n=1}^{\infty}(-1)^{n}e^{-2n\alpha}$. This expansion is absolutely converging wherever $|e^{-\alpha}|<1$, which is satisfied for the domain of integration provided $\zb>0$. Inserting the series expansion for $\tanh(\alpha)$ into equation \ref{eq:C_eqn} we obtain:

\begin{equation}
\int_{0}^{\infty}dz\frac{ze^{-z}}{\beta}\left[1-\frac{1}{\alpha}-\frac{2}{\alpha}\sum_{n=1}^{\infty}(-1)^{n}e^{-2n\alpha}\right]=\frac{1}{\beta}-\frac{2}{\beta^{3/2}}\Gamma(1/2)-\frac{4}{\beta^{3/2}}\int_{0}^{\infty}dzz^{-1/2}e^{-z}\left[\sum_{n=1}^{\infty}(-1)^{n}e^{-n\beta^{1/2}z^{3/2}}\right]
\end{equation}
Where $\Gamma(x)$ is the Gamma function. For the remaining integral terms on the right hand side, we expand $e^{-z}$ using Taylor series, $e^{-z}=\sum_{m=0}^{\infty}\frac{(-1)^{m}}{m!}z^{m}$. Once again this series is absolutely converging on the domain of integration.

\begin{equation}
-\frac{4}{\beta^{3/2}}\int_{0}^{\infty}dz\,z^{-1/2}e^{-z}\left[\sum_{n=1}^{\infty}(-1)^{n}e^{-n\beta^{1/2}z^{3/2}}\right]=-\frac{4}{\beta^{3/2}}\int_{0}^{\infty}dz\,z^{-1/2}\left[\sum_{m=0}^{\infty}\frac{(-1)^{m}}{m!}z^{m}\right]\left[\sum_{n=1}^{\infty}(-1)^{n}e^{-n\beta^{1/2}z^{3/2}}\right]\label{eq:product_of_sums}
\end{equation}

Because both infinite sums are absolutely converging on the domain of integration, their product in the right hand side of equation \ref{eq:product_of_sums} can be represented as a double summation:

\begin{equation}
-\frac{4}{\beta^{3/2}}\int_{0}^{\infty}dz\,z^{-1/2}\sum_{m=0}^{\infty}\sum_{n=1}^{\infty}\frac{(-1)^{m+n}}{m!}z^{m}e^{-n\beta^{1/2}z^{3/2}}\label{eq:double_sum}
\end{equation}

Next, we exchange the order of the integrand and the double summand. A proof for the validity of this step is provided in appendix \ref{sec:Proof-for-Equation}:

\begin{equation}
-\frac{4}{\beta^{3/2}}\sum_{m=0}^{\infty}\sum_{n=1}^{\infty}\frac{(-1)^{m+n}}{m!}\int_{0}^{\infty}dz\,z^{-1/2}z^{m}e^{-n\beta^{1/2}z^{3/2}}\label{eq:integral_sum_swap}
\end{equation}

Applying a change of variable, with $u=n\beta^{1/2}z^{3/2}$, equation \ref{eq:integral_sum_swap} becomes:

\begin{equation}
-\frac{4}{\beta^{3/2}}\sum_{m=0}^{\infty}\sum_{n=1}^{\infty}\frac{(-1)^{m}(-1)^{n}}{m!}\int_{0}^{\infty}du\left[\frac{2}{3n\beta^{1/2}}\right]\left[\frac{u}{n\beta^{1/2}}\right]^{2(m-1)/3}e^{-u}
\end{equation}

simplifying:

\begin{equation}
-\frac{8}{3\beta^{5/3}}\sum_{m=0}^{\infty}\sum_{n=1}^{\infty}\frac{(-1)^{m}(-1)^{n}}{m!}\frac{n^{-(1/3+2m/3)}}{\beta^{m/3}}\int_{0}^{\infty}du\,u^{2(m-1)/3}e^{-u}
\end{equation}

\begin{equation}
-\frac{8}{3\beta^{5/3}}\sum_{m=0}^{\infty}\sum_{n=1}^{\infty}\left[\Gamma(1/3+2m/3)\frac{(-1)^{m}}{m!\beta^{m/3}}\right](-1)^{n}n^{-(1/3+2m/3)}\label{eq:simplified_double_sum}
\end{equation}

The sum over $n$ in equation \ref{eq:simplified_double_sum} can be evaluated analytically for any value of $m$, giving:

\begin{equation}
-\frac{8}{3\beta^{5/3}}\sum_{m=0}^{\infty}a_{m}\frac{1}{\beta^{m/3}}\label{eq:final_series}
\end{equation}

\begin{equation}
a_{m}=\begin{cases}
(2^{2(1-m)/3}-1)\frac{(-1)^{m}}{m!}\zeta(1/3+2m/3)\Gamma(1/3+2m/3) & \text{if \ensuremath{m\neq1}}\\
\ln(2) & \text{if \ensuremath{m=1}}
\end{cases}
\end{equation}

Note that we are using here that $\sum_{n=1}^\infty (-1)^n/n^s = (2^{1-s}-1) \zeta(s)$ where $\zeta(s)$ is the classical Riemann Zeta function which for $s>1$ equals $\sum_{n=1}^\infty 1/n^s$ and which is defined for other complex $s$ by analytic continuation. The Riemann Zeta function should not be confused with the dimensionless breaking time $\zeta_B$. The series of equation \ref{eq:final_series} is absolutely convergent via the ratio test and because it is an alternating series the error $e_{M}$ of truncating after $M\gg1$ terms is bounded by $e_{M}\leq|a_{m}\beta^{-m/3}|$. Convergence should be fastest for large values of $\beta$ (fast-breaking systems, $\zb\ll1$) and slowest for small values of $\beta$ (slow breaking systems $\zb\gg1$ at low frequencies $\omega\trep\ll1$).

Consolidating to a final result for $G^{*}(\omega)$, the preceding analysis yields:

\begin{equation}
\frac{G^{*}(\omega)}{G_{e}}=i\omega\trep\left[\frac{1}{C(\zb,\omega)}-\frac{1}{\zb}\right]^{-1}
\end{equation}

\begin{equation}
C(\zb,\omega)=\frac{1}{\beta}-\frac{2}{\beta^{3/2}}\Gamma(1/2)-\frac{8}{3\beta^{5/3}}\sum_{m=0}^{\infty}a_{m}\frac{1}{\beta^{m/3}}\label{eq:solution}
\end{equation}

\begin{equation}
a_{m}=\begin{cases}
(-1)^{m}\frac{\Gamma(1/3+2m/3)}{m!}(2^{2(1-m)/3}-1)\zeta(1/3+2m/3) & \text{if \ensuremath{m\neq1}}\\
\ln(2) & \text{if \ensuremath{m=1}}
\end{cases}
\end{equation}

\begin{equation}
\beta=\zb^{-1}+i\omega\trep
\end{equation}

For readers that find the full analytic solution visually intimidating, keeping only the first three terms yields a good approximation for $\zb<0.001$.

\begin{equation}
C(\zb,\omega)\approx\frac{1}{\beta}-\frac{2\sqrt{\pi}}{\beta^{3/2}}\left[1-1.1522\frac{1}{\beta^{1/6}}\right]\label{eq:three_term}
\end{equation}

Unlike the single-mode Maxwell of equation \ref{eq:single_maxwell}, the three-term expansion in equation \ref{eq:three_term} is a valid description of tube-scale stress relaxation at all frequencies (i.e. not limited to $\omega\tb\ll1$) and leaves no undefined prefactors in the definition of the Maxwell time $\tau$. For many living polymer materials, the three-term expansion may be sufficient to specify both $\tb$ and $\trep$.

\section{\label{sec:Validation}Validation}

To evaluate the performance of truncated series solutions, numerical solutions to equation \ref{eq:C_eqn} were found via trapezoid quadrature; we use 100 log-spaced points on the interval $z\in[10^{-2},10]$, and results are converged to within the thickness of the lines shown. Series solutions following equation \ref{eq:solution} were calculated in Matlab with a trucation after $M$ terms. The native Matlab function for computing factorials and gamma functions is limited to arguments below 170, and larger factorials can be evaluated using variable-precision arithmetic or Stirling's approximation (accurate to at least four significant digits for the purposes of this work).

\begin{figure}
\begin{centering}
\includegraphics[width=0.5\columnwidth]{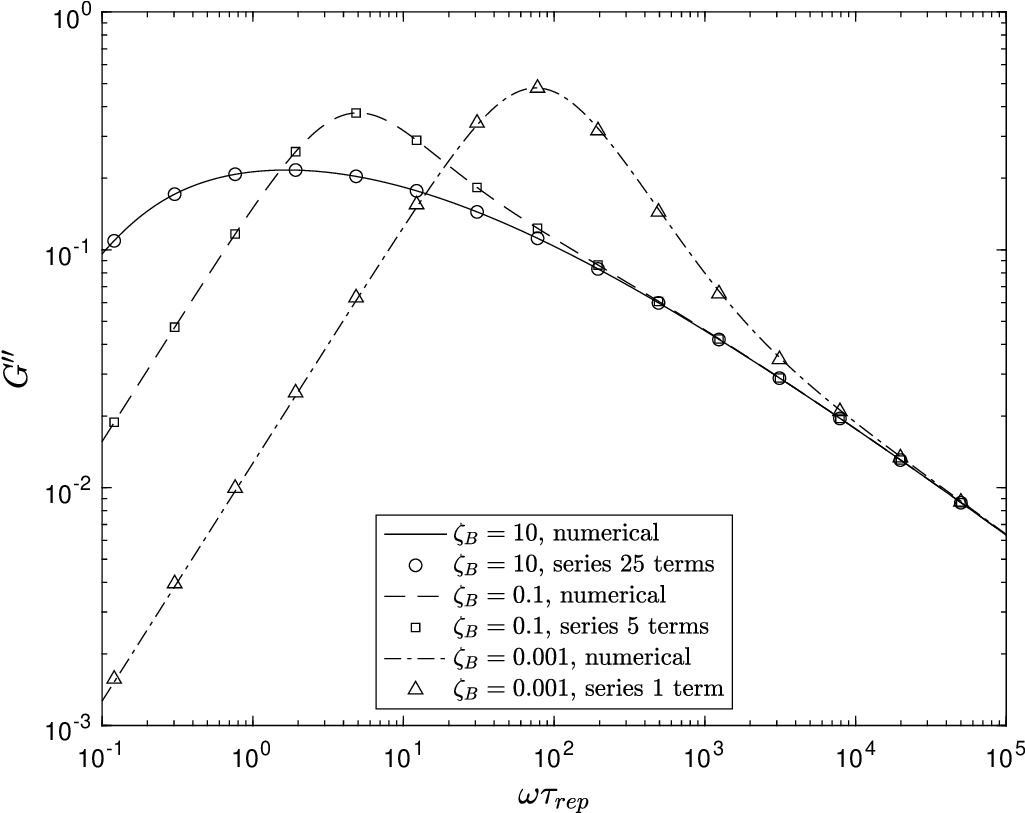}
\par\end{centering}
\caption{\label{fig:good_performance} Comparing converged numerical results from equation \ref{eq:C_eqn} (solid lines) against a truncated infinite series solution, equation \ref{eq:solution} (open symbols) for $\protect\zb=0.001,0.1,10$. The two methods agree to within the thickness of the line shown. The minimum number of terms needed to ensure such agreement at all frequencies increases with decreasing $\protect\zb$ as expected.}
\end{figure}

Figure \ref{fig:good_performance} demonstrates the performance of the truncated series solutions by comparing against converged numerical solutions for $\zeta_{B}$ ranging from $10^{-3}$ to $10$, covering both fast-breaking $\zb\ll1$ and slow breaking $\zb\gg1$ behaviors. For each curve shown the series solutions includes approximately the minimum number of terms to achieve convergence (within the thickness of the line) for all frequencies. In the fast-breaking limit, $\zeta_{B}\ll1$, the summation solution converges quickly to the numerical solution; for $\zeta_{B}<10^{-3}$ only the first term of the series is necessary (c.f. equation \ref{eq:three_term}). Increasing $\zeta_{B}$ toward the slow-breaking limit requires an increased number of terms in the summation to avoid deviation from the numerical solution at low frequencies.

\begin{figure}
\begin{centering}
\includegraphics[width=0.5\columnwidth]{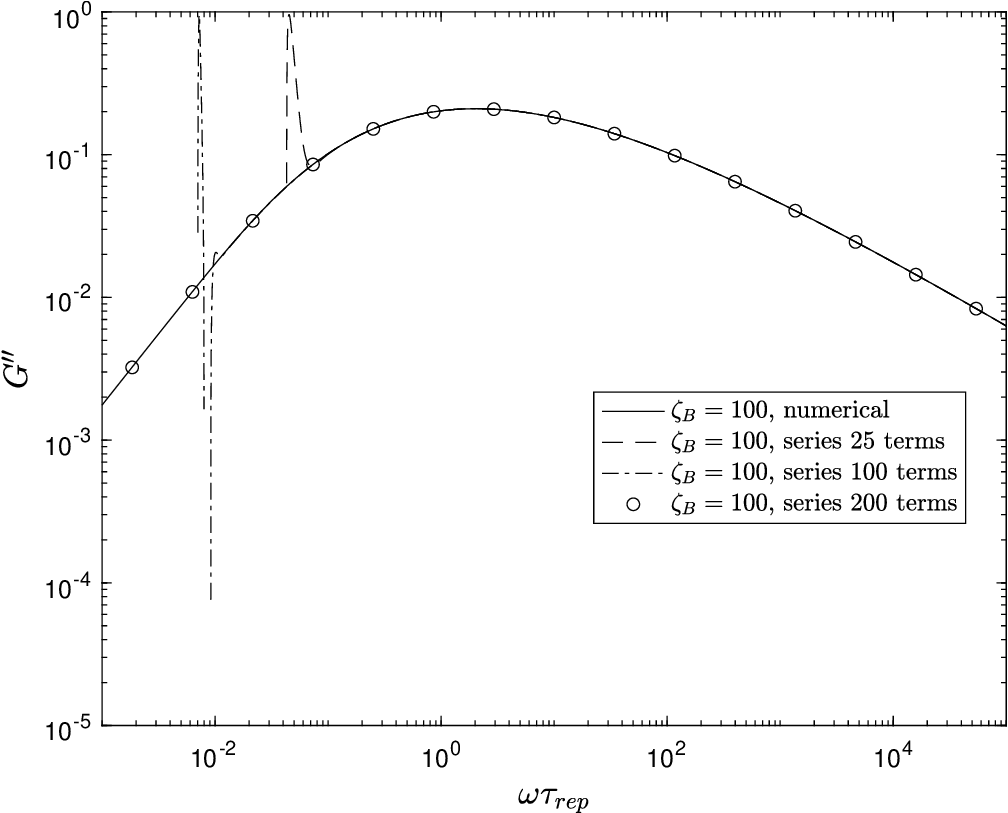}
\par\end{centering}
\caption{\label{fig:bad_performance}Comparing converged results from equation \ref{eq:C_eqn} (solid line) against a truncated infinite series solution, equation \ref{eq:solution} for $\protect\zb=100$ and thresholds for truncation, $M=25,100,200$. When too few terms are included (broken lines), the rheological predictions are pathological, but good agreement is found after including a sufficient number of terms (open symbols).}
\end{figure}

Finally, Figure \ref{fig:bad_performance} demonstrates a failure of the truncated series solutions when the series is trucated too soon for large values of $\zb$. The failure of the series solution appears as a discontinuity in the complex modulus; the frequency about which this discontinuity is centered moves toward lower frequencies as the number of terms $M$ in the series increases. Empirically, we have observed that once the pathological behavior is pushed to frequencies below $\omega\trep\sim1/\zb$, the divergence disappears for only modestly increased $M$. We also observe that for larger values of $\zb$ within the range considered here, the number of modes $M$ for good convergence scales approximately as $M\sim\zeta_{B}$ for $\zeta_{B}\gg1$.

Overall, the results shown in Figures \ref{fig:good_performance} and \ref{fig:bad_performance} confirm that our series solution of equation \ref{eq:solution} is valid for any finite value of $\zb$. The best convergence is found for fast-breaking systems, $\zb\ll1$, and the three-term approximation of equation \ref{eq:three_term} is sufficient for $\zb<10^{-3}.$

\section{\label{sec:Conclusions}Conclusions}

In this work, we have produced an analytic series solution for the shuffling model of well-entangled living polymers. The shuffling model is a variation of the classic Poisson renewal model, reframed as a differential constitutive equation and updated for a more physically realistic length-independent ``shuffling'' time \cite{peterson2023wormlike}. The analytic series solution was found by expanding the arguments of an integral into (absolutely converging) Taylor series, and then exchanging the order of integration and summation. Following a change of variable, the integral was found to have a closed-form solution, and the double summation could be collapsed to a single summation. The final result was an analytic solution for the complex modulus $G^{*}$ in terms of a single infinite series expansion. The infinite series expansion was found to converge for all values of $\omega\trep$ and all values of $\zb$, though convergence was fastest for small values of $\zb$.

Compared to other methods of describing living polymer rheology, the analytic solution shown here will generally be faster and easier to implement. It is worth noting, however, that our series solution does not include the effects of contour length fluctuation or thermal constraint release (e.g. double reptation), which makes it less useful than other methods for quantitative parameter evaluation. Even so, applicaitons that require more advanced/complete models may still find our analytic solution useful as a means of quickly arriving at good initial estimates of $\tb$, $\trep$, and $G_{e}$ for example.

In future work, it would be interesting to see if the series solution \ref{eq:solution} could be found from higher-order asymptotic methods for partial differential equations (i.e. boundary layer theory) to equation \ref{eq:shuffling}, as was partially-explored in a previous study \cite{peterson2020full}. Additionally, an analytic solution for $G^{*}(\omega)$ suggests an opportunity to learn more about the continuous/discrete relaxation spectra of the shuffling model. Overall, whereas the shuffling model is limited by its phenomenological basis of rearrangement, knowing the underlying structure of its solution in analytic form should yield transferrable insights for real living polymer systems.

\appendix

\section{\label{sec:Proof-for-Equation}Proof for Equation \ref{eq:integral_sum_swap}}

In equation \ref{eq:integral_sum_swap}, we claim that:
\begin{equation}
\int_{0}^{\infty}dz\,z^{-1/2}\sum_{m=0}^{\infty}\sum_{n=1}^{\infty}\frac{(-1)^{m+n}}{m!}z^{m}e^{-n\beta^{1/2}z^{3/2}}=\sum_{m=0}^{\infty}\sum_{n=1}^{\infty}\frac{(-1)^{m+n}}{m!}\int_{0}^{\infty}dz\,z^{m-1/2}e^{-n\beta^{1/2}z^{3/2}}.
\end{equation}

Interchange of an integral with a sum is allowed if there are only finitely many terms in the sum (and all integrals are finite), but interchange of an integral with a sum of infinitely many terms must be justified more carefully. A proof of this result is given in three steps:

\textbf{Step 1:} Divide the integral into 3 different integrals to seperate the $m=0$ and $m=1$ sums from the sum over $m\geq2$.

\begin{align}
\int_{0}^{\infty}dz\,z^{-1/2}\sum_{m=0}^{\infty}\sum_{n=1}^{\infty}\frac{(-1)^{m+n}}{m!}z^{m}e^{-n\beta^{1/2}z^{3/2}} & =\int_{0}^{\infty}dz\,z^{-1/2}\sum_{n=1}^{\infty}(-1)^{n}e^{-n\beta^{1/2}z^{3/2}}\nonumber \\
 & \qquad+\int_{0}^{\infty}dz\,z^{-1/2}\sum_{n=1}^{\infty}(-1)^{n+1}ze^{-n\beta^{1/2}z^{3/2}}\nonumber \\
 & \qquad+\int_{0}^{\infty}dz\,z^{-1/2}\sum_{m=2}^{\infty}\sum_{n=1}^{\infty}\frac{(-1)^{m+n}}{m!}z^{m}e^{-n\beta^{1/2}z^{3/2}}
\end{align}

\textbf{Step 2:} Justify interchanging the integral and the infinite sum over $n$ for the $m=0$ term. That is, we wish to show that
\begin{align*}
\int_{0}^{\infty}dz\,z^{-1/2}\sum_{n=1}^{\infty}(-1)^{n}e^{-n\beta^{1/2}z^{3/2}} & =\sum_{n=1}^{\infty}(-1)^{n}\int_{0}^{\infty}dz\,z^{-1/2}e^{-n\beta^{1/2}z^{3/2}}.
\end{align*}

To justify the interchange of the sum and integral, first note that the interchange would be allowed if the sum on the inside were only over finitely many $n$. That is, for any fixed $N<\infty$ we have 

\begin{equation}
\int_{0}^{\infty}dz\,z^{-1/2}\sum_{n=1}^{N-1}(-1)^{n}e^{-n\beta^{1/2}z^{3/2}}=\sum_{n=1}^{N-1}\int_{0}^{\infty}dz\,z^{-1/2}(-1)^{n}e^{-n\beta^{1/2}z^{3/2}}.
\end{equation}

Next, to compare the integral of the infinite sum with the integral of the finite sum, note that since the sum is an alternating series with terms decreasing in absolute value we have:

\begin{equation}
\left|z^{-1/2}\sum_{n=1}^{\infty}(-1)^{n}e^{-n\beta^{1/2}z^{3/2}}-z^{-1/2}\sum_{n=1}^{N-1}(-1)^{n}e^{-n\beta^{1/2}z^{3/2}}\right|\leq z^{-1/2}|e^{-N\beta^{1/2}z^{3/2}}|=z^{-1/2}e^{-rNz^{3/2}},
\end{equation}

where $r=\Re(\beta^{1/2})>0$ is the real part of $\beta^{1/2}$ (note that we are using here that $r=\Re(\beta^{1/2})>0$ since $\Im(\beta)\neq0$). Applying this inequality to the integrals we find:

\[
\left|\int_{0}^{\infty}dz\,z^{-1/2}\sum_{n=1}^{\infty}(-1)^{n}e^{-n\beta^{1/2}z^{3/2}}-\sum_{n=1}^{N-1}\int_{0}^{\infty}dz\,z^{-1/2}(-1)^{n}e^{-n\beta^{1/2}z^{3/2}}\right|=
\]

\begin{equation}
\left|\int_{0}^{\infty}dz\,z^{-1/2}\sum_{n=1}^{\infty}(-1)^{n}e^{-n\beta^{1/2}z^{3/2}}-\int_{0}^{\infty}dz\,z^{-1/2}\sum_{n=1}^{N-1}(-1)^{n}e^{-n\beta^{1/2}z^{3/2}}\right|\leq\int_{0}^{\infty}dz\,z^{-1/2}e^{-rNz^{3/2}}.
\end{equation}

Since the Dominated Convergence Theorem implies that this last integral converges to zero as $N\to\infty$ we can conclude that:

\begin{align}
\int_{0}^{\infty}dz\,z^{-1/2}\sum_{n=1}^{\infty}(-1)^{n}e^{-n\beta^{1/2}z^{3/2}} & =\lim_{N\to\infty}\sum_{n=1}^{N-1}\int_{0}^{\infty}dz\,z^{-1/2}(-1)^{n}e^{-n\beta^{1/2}z^{3/2}}\nonumber \\
 & =\sum_{n=1}^{\infty}\int_{0}^{\infty}dz\,z^{-1/2}(-1)^{n}e^{-n\beta^{1/2}z^{3/2}}.
\end{align}

\textbf{Step 3:} Justify interchanging the integral and the infinite sum for $m=1.$ That is, we wish to show that 

\begin{align*}
\int_{0}^{\infty}dz\,z^{-1/2}\sum_{n=1}^{\infty}(-1)^{n+1}ze^{-n\beta^{1/2}z^{3/2}} & =\sum_{n=1}^{\infty}\int_{0}^{\infty}dz\,z^{-1/2}(-1)^{n+1}ze^{-n\beta^{1/2}z^{3/2}}.
\end{align*}

This can be justified exactly as in the $m=0$ case.

\textbf{Step 4:} Justify interchanging the integral with the infinite sum. That is, justify that 
\[
\int_{0}^{\infty}dz\,z^{-1/2}\sum_{m=2}^{\infty}\sum_{n=1}^{\infty}\frac{(-1)^{m+n}}{m!}z^{m}e^{-n\beta^{1/2}z^{3/2}}=\sum_{m=2}^{\infty}\sum_{n=1}^{\infty}\int_{0}^{\infty}dz\,z^{-1/2}\frac{(-1)^{m+n}}{m!}z^{m}e^{-n\beta^{1/2}z^{3/2}}.
\]
This will be justified by Fubini's Theorem (viewing the sums as integrals with respect to counting measure) if we can show that the sums of the integrals are still finite when we take the absolute value of the integrands. That is, we need to show 
\begin{equation}
\sum_{m=2}^{\infty}\sum_{n=1}^{\infty}\int_{0}^{\infty}dz\,z^{-1/2}\frac{1}{m!}z^{m}e^{-nrz^{3/2}}<\infty,\label{suff}
\end{equation}
where again $r=\Re(\beta^{1/2})>0$. To show this, use a change of variables $u=nrz^{3/2}$ to evaluate the inner integral.

\begin{align}
\sum_{m=2}^{\infty}\sum_{n=1}^{\infty}\int_{0}^{\infty}dz\,z^{-1/2}\frac{1}{m!}z^{m}e^{-nrz^{3/2}} & =\sum_{m=2}^{\infty}\frac{1}{m!}\sum_{n=1}^{\infty}\int_{0}^{\infty}du\,\left(\frac{2}{3nr}\right)\left(\frac{u}{nr}\right)^{2(m-1)/3}e^{-u}\nonumber \\
 & =\frac{2}{3}r^{-1/3}\sum_{m=2}^{\infty}\frac{1}{m!}r^{-2m/3}\sum_{n=1}^{\infty}n^{-(1/3+2m/3)}\Gamma(1/3+2m/3)\nonumber \\
 & =\frac{2}{3}r^{-1/3}\sum_{m=2}^{\infty}\frac{1}{m!}r^{-2m/3}\zeta(1/3+2m/3)\Gamma(1/3+2m/3),
\end{align}

and the last sum above is finite by the ratio test. Thus, we have shown equation (\ref{suff}). Taken together, these steps confirm that the order of the integral and sum can be exchanged as in equation \ref{eq:integral_sum_swap} of our derivation in section \ref{sec:Derivation}.

\printbibliography

\end{document}